\documentclass[amsmath,amssymb,aps,prl,oneocolumn,superscriptaddress,showpacs,floatfix,notitlepage]{revtex4-1}

\usepackage{graphicx}
\usepackage{color}
\usepackage{epstopdf}
\usepackage{braket}
\usepackage{ulem}

\begin{document} 
\title{Atomic-scale spin sensing with a single-molecule at the apex of a scanning tunneling microscope}
\author{B. Verlhac}
\author{N. Bachellier}
\author{L. Garnier}
\author{M. Ormaza}
\affiliation{Universit\'{e} de Strasbourg, CNRS, IPCMS, UMR 7504, F-67000 Strasbourg, France}
\author{P. Abufager}
\affiliation{Instituto de F\'{i}sica de Rosario, Consejo Nacional de Investigaciones Cient\'{i}ficas y T\'ecnicas (CONICET) and Universidad Nacional de Rosario, Av. Pellegrini 250 (2000) Rosario, Argentina}
\author{R. Robles}
\affiliation{Catalan Institute of Nanoscience and Nanotechnology (ICN2), CSIC and BIST, Campus UAB, Bellaterra, 08193 Barcelona, Spain}
\author{M.-L. Bocquet}
\affiliation{PASTEUR, D\'epartement de Chimie, Ecole Normale Sup\'erieure, PSL Research University, Sorbonne Universit\'es, UPMC Univ. Paris 06, CNRS, 75005 Paris, France}
\author{M. Ternes}
\affiliation{Institute of Physics II B, RWTH Aachen University, 52074 
Aachen, Germany}
\affiliation{Peter Gr\"unberg Institut (PGI-3), Forschungszentrum J\"ulich, 
52425 J\"ulich, Germany}
\author{N. Lorente}
\affiliation{Donostia International Physics Center (DIPC), 20018 Donostia-San Sebasti\'an, Spain}
\affiliation{Centro de F{\'{\i}}sica de Materiales (CFM), 20018 Donostia-San Sebasti\'an, Spain}
\author{L. Limot}
\email{limot@ipcms.unistra.fr}
\affiliation{Universit\'{e} de Strasbourg, CNRS, IPCMS, UMR 7504, F-67000 Strasbourg, France}

\date{\today}

\begin{abstract}
\textbf{Recent advances in scanning probe techniques rely on the chemical functionalization of the probe-tip termination by a single molecule. The success of this approach opens the tantalizing prospect of introducing spin sensitivity through the functionalization by a magnetic molecule. Here, we use a nickelocene-terminated tip (Nc-tip), which offers the possibility of producing spin excitations on the tip apex of a scanning tunneling microscope (STM). We show that when the Nc-tip is a hundred pm away from point contact with a surface-supported object, magnetic effects may be probed through changes in the spin excitation spectrum of nickelocene. We use this detection scheme to simultaneously determine the exchange field and the spin polarization of the sample with atomic-scale resolution. Our findings demonstrate that the Nc-tip is a powerful probe for investigating surface magnetism with STM, from single magnetic atoms to surfaces.}
\end{abstract}

\maketitle 

In a conventional STM setup the magnetic ground state of an isolated object \textemdash atom or molecule\textemdash is inferred by collecting spin-related fingerprints in the conductance measured with a metallic tip. These isolated objects may also serve as spin detectors when they are controllably moved on the surface with the help of the tip within their local magnetic environment. The magnetic ground state is in fact prone to change in the presence of a magnetic coupling. Exchange and surface-mediated Ruderman-Kittel-Kasuya-Yosida interactions have been spatially mapped in this way by monitoring the zero-bias peak in the differential conductance ($dI/dV$) associated to the Kondo effect~\cite{Otte2009,Tsukahara2011,Neel2011,Fu2012}, the tunneling magnetoresistance~\cite{Meier2008,Khajetoorians2011a,Khajetoorians2011b}, the spin excitation spectra and spin relaxation times~\cite{Hirjibehedin2006,Chen2008,Loth2010,Yan2017}. Recently, also dipolar and hyperfine interactions have been observed via electrically-driven spin resonances~\cite{Natterer2017,Willke2018}. 

\begin{figure}
  \includegraphics[width=0.95\textwidth,clip]{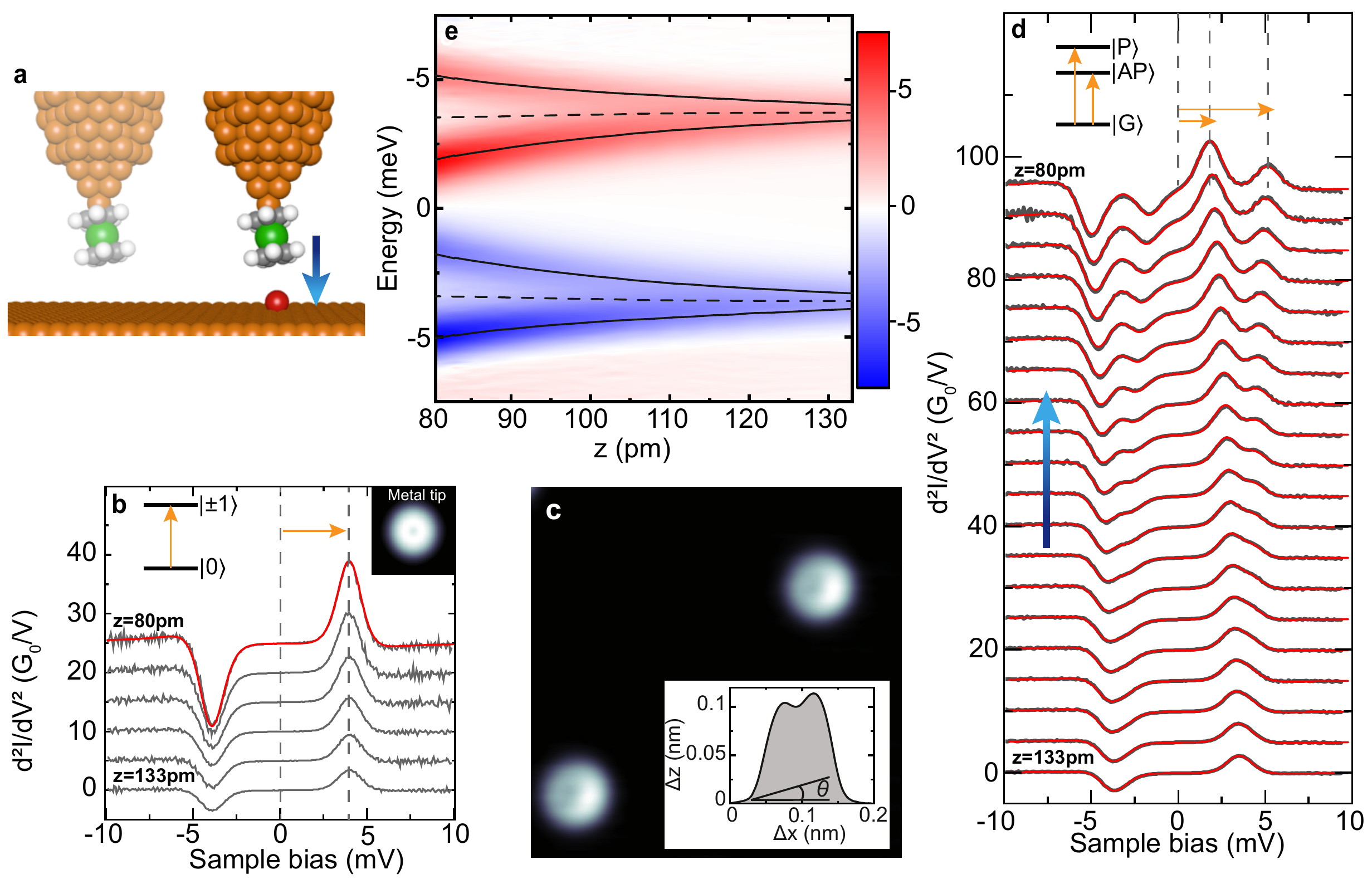}
  \caption{\textbf{Spin excitation spectra above the Cu surface and a single Fe atom.} (a) Schematic view of the tunnel junction. Atom colors: Cu (orange), C (grey), H (white), Ni (green), Fe (red). (b) $d^2I/dV^2$ spectra acquired above a surface atom of Cu(100) at a distance between $z=133$ pm and $z=80$~pm. The solid red line is a fit based on a dynamical scattering model~\cite{Ternes2015}, and yields an axial magnetic anisotropy of $D=3.5$~meV, a coupling between the localized Nc spin and the tip electrons of $J_0\rho_0=-0.08$, and a spin-conserving potential scattering of $U=0.02$. Inset: Image of Nc on Cu(100) acquired with a copper-coated tip apex ($V=20$~mV, $I=100$~pA, size: $2\times2$~nm$^2$). (c) Image of Fe atoms on Cu(100) ($V=-15$~mV, $I=30$~pA, size: $5\times5$~nm$^2$), and corresponding line profile of one atom (Inset) revealing the presence of a tilted Nc at the tip apex. The tilt angle $\theta$ is estimated via the height difference between the left an right protrusion of the line profile. (d) $d^2I/dV^2$ spectra acquired with the tip positioned above the high-intensity side of the ring-shaped Fe atom. Shorter tip distances than $z=80$~pm result in the transfer of Nc atop the Fe atom (Figs. S1b and S1c). (e) Peak and dip energy positions extracted from the spectra of panel (d). The color scale flanking the panel corresponds to the $d^2I/dV^2$ amplitudes and is given in units of $G_0/$V. For clarity, the spectra in (b) and (d) are shifted vertically from one another by $5\times G_0$/V ($G_0=2e^{2}/h$: quantum of conductance).
\label{fig1}}
\end{figure}

Having a well-calibrated sensor attached to the tip apex would provide a more advantageous setup since the tip can be freely positioned above a target object on the surface. This detection scheme eliminates surface-mediated interactions and benefits from the vertical-displacement sensitivity of the STM as the sensor-object distance is no longer imposed by the surface corrugation. Probing a magnetic exchange interaction across a vacuum gap is however experimentally demanding~\cite{Kaiser2007,Bork2011,Schmidt2011,Yan2014,Choi2016,Muenks2017,Schmidt2011} as scanning probe techniques suffer, unfortunately, from the poor structural and magnetic knowledge of the tip apex. To overcome this limitation, we introduce spin sensitivity by functionalazing the tip apex with a single magnetic molecule. Such a strategy has proven successful for collecting chemical and structural information on surface-supported objects otherwise inaccessible with a metallic tip~\cite{Gross2009,Weiss2010,Chiang2014,Wagner2015,Guo2016,Monig2018}. We use here a tip decorated by a spin $S=1$ nickelocene molecule (Fig.~\ref{fig1}a)~\cite{Ormaza2017a,Ormaza2017b}, which comprises a Ni atom sandwiched between two C$_5$H$_5$ cyclopentadienyl rings, and accurately capture the junction geometry through the comparison to first-principles calculations.

We prepare the Nc-tip with atomic control. We first perform soft tip-surface indentations into our pristine working surfaces, either Cu(100) or Cu(111) (Supplemental Sec. I), to ensure a mono-atomically sharp Cu apex. Nickelocene is then imaged as a ring (inset in Fig.~\ref{fig1}b), since the molecule adsorbs on copper with one cyclopentadienyl bound to the surface while the other is exposed to vacuum~\cite{Bachellier2016}. After transferring the Nc molecule from the surface to the tip \textemdash details of the molecule transfer to the tip can be found in~\cite{Ormaza2017b}, the Nc-tip is characterized by spectral features found in the second derivative, $d^2I/dV^2$, of the current $I$ with respect to the bias $V$ measured at a constant distance $z$ between tip and the pristine surface. Figure~\ref{fig1}b presents a set of $d^2I/dV^2$ spectra recorded a different $z$ heights above Cu(100). We calibrate $z$ by performing controlled tip contacts to the surface tracking the current $I$ and defining $z=0$ as the distance where the transition from the tunneling to contact regime occurs (Supplemental Sec.~II and Fig.~S1a). The spectra vary with $z$ only in amplitude and are dominated by a peak at positive and a dip at negative $V$ at energies symmetric to zero. These peaks and dips correspond to inelastic tunneling events in which tunneling electrons excite the Nc from its magnetic ground state $\ket{M=0}$, with $M$ as the magnetic quantum number projected onto the axis perpendicular to the rings of the molecule, to one of the two degenerate excited states $\ket{M=\pm1}$. These states are at a higher energy $D=(3.5\pm0.1)$~meV relative to the ground state \cite{Ormaza2017a}, where $D$ is the axial magnetic anisotropy [see Eq.~(\ref{ham})]. The inelastic conductance is nearly one order of magnitude higher than the elastic conductance (Fig.~S2), highlighting that the spin of nickelocene is well preserved from scattering events with itinerant electrons of the metal~\cite{Ormaza2017a,Rubio2018}. This behavior is remarkable, differentiating Nc from other single objects, which instead require a thin insulating spacer between them and the metal surface~\cite{Heinrich2004,Hirjibehedin2006,Rau2014} or a superconductor~\cite{Heinrich2013} to preserve their quantum nature. 

We use the Nc-tip to probe surface magnetism through changes in the spin excitation spectrum as we first demonstrate by approaching a single magnetic Fe atom adsorbed on Cu(100) (Fig.~\ref{fig1}a). Iron atoms on the surface (Supplemental Sec. I) protrude by $115$~pm and are imaged with the Nc-tip as rings with an asymmetric apparent height (Fig.~\ref{fig1}c). The structure observed in the image reflects the presence of Nc on the tip apex and is a consequence of the tilted adsorption geometry of the molecule. Indeed, density functional theory (DFT) calculations show that Nc bonds to the the tip-apex atom through two C atoms of the cyclopentadienyl ring~\cite{Ormaza2017b}. The tilt angle can be estimated through the line profile of the Fe atom (Inset of Fig.~\ref{fig1}c) revealing for the tips used typically angles $\lesssim15^{\circ}$ relative to the surface normal. Figure~\ref{fig1}d presents a set of $d^2I/dV^2$ spectra recorded at different $z$ heights above the Fe atom. At $z=133$~pm, the $d^2I/dV^2$ spectrum is indistinguishable from the one acquired above the bare Cu(100) (Fig.~\ref{fig1}b), which indicates that in this low-energy range Fe is spectroscopically dark, \textit{i.e.}\ its contribution to the spin excitation spectrum is negligible. However, for smaller $z$ heights we observe a splitting of the peak and the dip which becomes increasingly stronger as the tip-Fe distance decreases (Fig.~\ref{fig1}e), the average position of the spin-split peaks and dips varying at most by $0.2$~meV with distance [(dashed line in Fig.~\ref{fig1}e)]. So far we tested several Nc-tips with tilt angles ranging from $5^{\circ}$ to $15^{\circ}$, and all showed similar behavior \textemdash the distances for a given splitting changing only by $\pm10$~pm. Apart from the splitting we also observe a striking intensity asymmetry of the split spectral features. While the amplitude of the positive peak and negative dip are identical for the Nc-tip probed against the bare surface (Fig.~\ref{fig1}b), here we find that at positive bias the energetically lower excitation has higher peak amplitude than the energetically higher excitation, while at negative bias the dips show opposite behavior compared to the peaks. 

\begin{figure}
  \includegraphics[width=0.80\textwidth,clip]{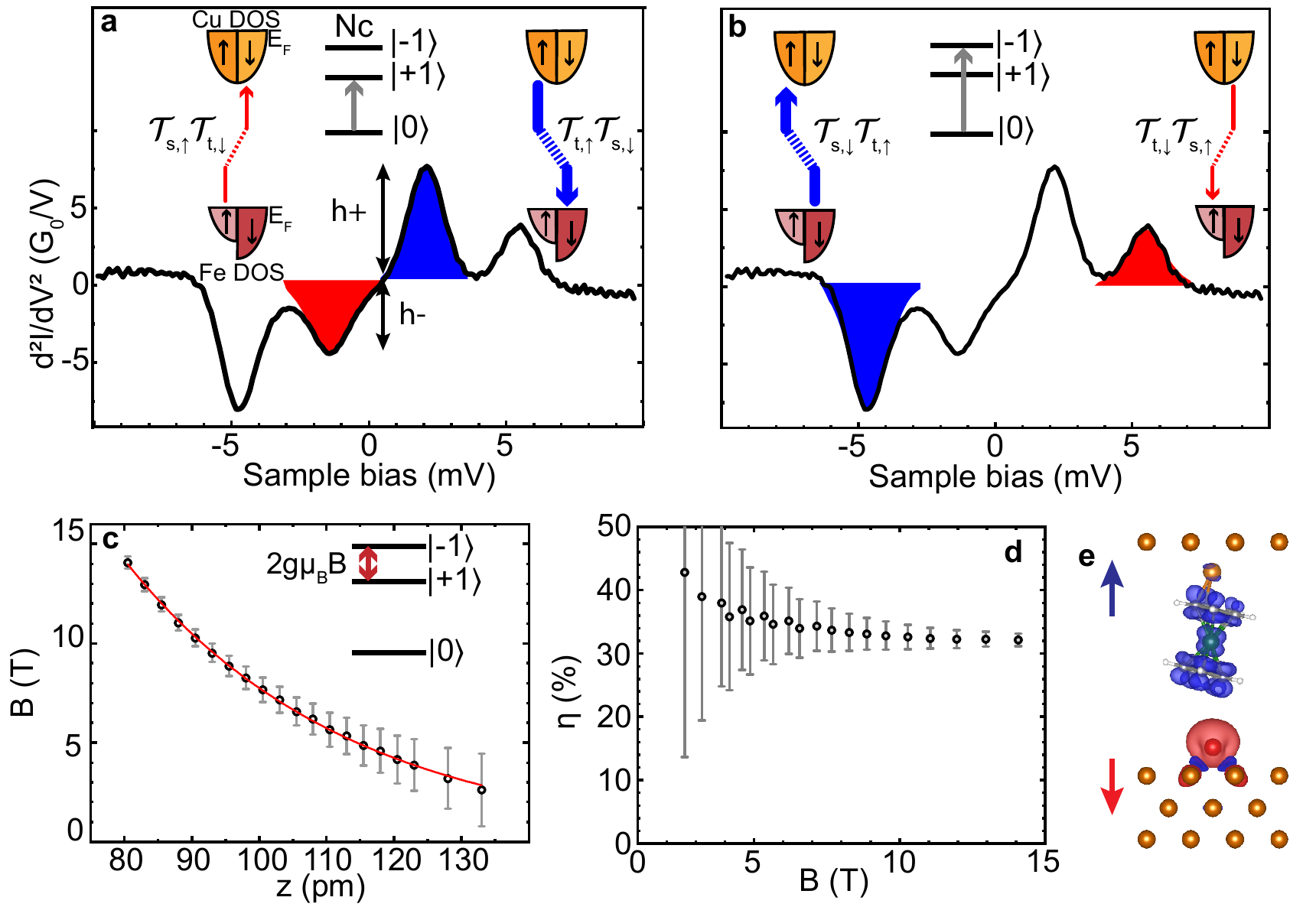}
  \caption{\textbf{Inelastic tunneling and magnetic coupling measured above a Fe atom.} Panels (a) and (b) sketch the mechanism leading to the bias asymmetry in the $d^2I/dV^2$ spectra acquired above Fe. (a) At negative bias, inelastic electrons tunnel from the spin-up states of the Fe atom to the spin-down states of the copper tip with transmission proportional to $\mathcal{T}_{\text{t},\downarrow}\mathcal{T}_{\text{s},\uparrow}$, leading to a dip in the $d^2I/dV^2$ spectrum. At positive bias, instead, the junction polarity is reversed and inelastic electrons tunnel from the spin-up states of the copper tip to the spin-down states of the Fe atom with transmission proportional to $\mathcal{T}_{\text{t},\uparrow}\mathcal{T}_{\text{s},\downarrow}$, leading to a peak in the $d^2I/dV^2$ spectrum. During the tunneling process, the electrons excite Nc from its ground to its first excited state. The weaker amplitude for the dip compared to the peak reflects the difference in the transmission ($\mathcal{T}_{\text{t},\uparrow}\mathcal{T}_{\text{s},\downarrow}\neq\mathcal{T}_{\text{t},\downarrow}\mathcal{T}_{\text{s},\uparrow}$). DOS: density of states. (b) Same mechanism as (a) but for inelastic tunnel electrons exciting Nc from its ground state to its second excited state. (c) Exchange field $B$ and (d) spin-asymmetry $\eta$ extracted from the spectra of Fig.~\ref{fig1}d using Eq.~(\ref{ham}). (e) DFT calculated configuration of the tunnel junction for a distance of $100$ pm with isosurface of the spin density (antiferromagnetic coupling).
\label{fig2}}
\end{figure}

The splitting and asymmetry of the line shape observed above Fe have magnetic origin. To rationalize these observations, we assign the $z$-axis as the out-of-surface direction neglecting the small tilt of the molecule and employ a spin Hamiltonian that includes the magnetic anisotropy of the Nc molecule and of the Fe atom on copper ($D_\text{Fe}$)~\cite{Khajetoorians2011b,Pacchioni2015}
\begin{equation}
H=D\hat{S}_{z}^2+D_\text{Fe}\hat{S}_{z, \text{Fe}}^2-J \hat{S}_{z} \cdot \hat{S}_{z,\text{Fe}},
\label{eq1}
\end{equation}
where $J$ is a Ising-like exchange coupling restraining the Nc-Fe magnetic interaction along the $z$-axis (Supplemental Sec.~III, Fig.~S3). Within this framework, the ground state of the combined system is a doublet $\ket{\text{G}}=\ket{0, \Uparrow}$ and $\ket{0, \Downarrow}$ where the energetically lowest states of the Fe spin are noted $\Uparrow$ and $\Downarrow$. The exchange interaction lifts the degeneracy between the two excited doublets $\ket{\text{AP}}$ and $\ket{\text{P}}$ of the coupled spin system and causes the line shape to split apart. As the exchange interaction is antiferromagnetic ($J<0$, see below) and the Fe is spectroscopically dark, the lowest excited state doublet is $\ket{\text{AP}}=\ket{-1, \Uparrow}$ and $\ket{+1, \Downarrow}$ and corresponds to an antiferromagnetic configuration where the Fe spin is anti-aligned with the Nc spin. The higher excited state doublet corresponds to the ferromagnetic configuration, $\ket{\text{P}}=\ket{+1,\Uparrow}$ and $\ket{-1, \Downarrow}$. Note that for this derivation neither the spin magnitude of the Fe atom nor the sign of $D_{\text{Fe}}$ has to be explicitly set as long as the ground state is a doublet.

To simplify the discussion, it is however preferable to express the spin Hamiltonian of Eq.~(\ref{eq1}) with an effective Zeeman term consisting of the gyromagnetic factor ($g$), the Bohr magnetron ($\mu_\text{B}$), and the exchange field ($B$) produced by the Fe atom and acting along the $z$-axis of the Nc molecule:
\begin{equation}
\hat{H}=D \hat{S_z}^2-g\mu_\text{B}B\hat{S_z}.
\label{ham}
\end{equation}
This expression presents also the advantage of providing a common framework for describing the spin systems investigated in the present study. Within mean-field theory $B= J \langle S_\text{Fe}\rangle/g\mu_\text{B}$, where $\langle S_\text{Fe}\rangle$ is the effective spin of Fe on the Cu(100) surface. In the following, for clarity we restrain the analysis to a $\Downarrow$ Fe spin, without loss of generality (Supplemental Sec. III, Fig.~S4). The exchange field, within this viewpoint, causes a Zeeman splitting of the line shape into the two excited states $\ket{+1}$ and $\ket{-1}$ of Nc that are located at low and high energy, respectively. The bias asymmetry in the peaks and dips reflects instead a spin imbalance in the tunneling current~\cite{Balashov2006,Loth2010,Delgado2010,Novaes2010}. This mechanism is sketched in Figs.~\ref{fig2}a and \ref{fig2}b. It is qualitatively similar to conventional spin-polarized STM \cite{Meier2008} and, more generally, to spin valves or to Kondo systems coupled to magnetic electrodes~\cite{Pasupathy04,Fu2012,Bergmann2015,Choi2016}. As illustrated in Fig.~\ref{fig2}a, the excitation of Nc from the ground state to its first excited state $\ket{+1}$ requires a change in spin angular momentum of $\delta M=+1$.  Since the total angular momentum has to be conserved it can only be induced by electrons that compensate for this moment by flipping their spin direction during the tunneling process from $\ket{\uparrow}$ to $\ket{\downarrow}$. At negative bias, the tunneling process may be viewed as a $\ket{\uparrow}$ electron hopping from the substrate (noted $\text{s}$) into the molecular orbital, while a $\ket{\downarrow}$ electron is emitted from the molecular orbital towards the tip (noted $\text{t}$). Using second-order perturbation theory, we assign an amplitude $\mathcal{T}_{\text{s},\uparrow}$ to the first process and $\mathcal{T}_{\text{t},\downarrow}$ to the second process, so that the transmission from sample to tip ($\text{s}\rightarrow \text{t}$) is proportional to $\mathcal{T}_{\text{s},\uparrow}\mathcal{T}_{\text{t},\downarrow}$. At positive bias, the tunneling direction reverts and the transmission from tip to sample ($\text{t}\rightarrow \text{s}$) is proportional to $\mathcal{T}_{\text{t},\uparrow}\mathcal{T}_{\text{s},\downarrow}$. The relative height of the dip at low negative voltage ($h_{-}$) compared to the peak at low positive voltage ($h_{+}$) yields a quantitative measure of the spin asymmetry $\eta=(h_{+}-h_{-})/(h_{+}+h_{-})$, with $h_{-}\propto \mathcal{T}_{\text{s},\uparrow}\mathcal{T}_{\text{t},\downarrow}$ and $h_{+}\propto \mathcal{T}_{\text{t},\uparrow}\mathcal{T}_{\text{s},\downarrow}$. The excitation of Nc from the ground state to its second excited state $\ket{-1}$ requires instead electrons starting in a $\ket{\downarrow}$ state and ending in a $\ket{\uparrow}$ state (Fig.~\ref{fig2}b), resulting in a spin asymmetry of $-\eta$. Due to the mixing of these spin-dependent amplitudes, the observed spin asymmetry  differs from the usual spin polarization measured with elastic electrons.

\begin{figure}
  \includegraphics[width=0.95\textwidth,clip]{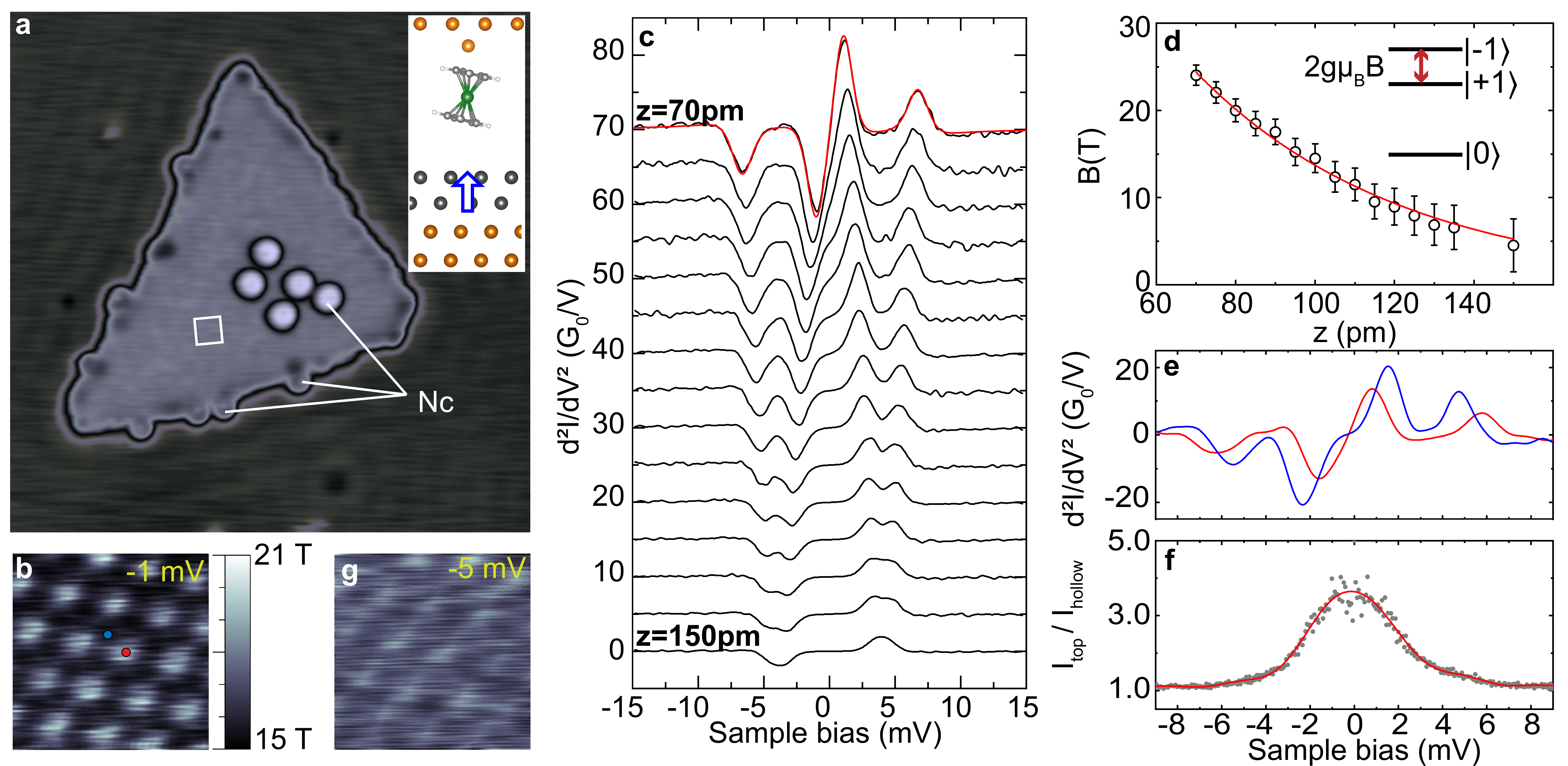}
  \caption{\textbf{Spin excitation spectra above a cobalt surface.} (a) Cobalt island on Cu(111) decorated with Nc molecules ($V=-50$~mV, $I=20$~pA, size: $25\times25$~nm$^2$). The white lines highlight the presence of nickelocene. The white square corresponds to the area investigated in panels (b) and (f). Inset: Schematic view of a Nc-tip above a cobalt island. The arrow indicates the out-of-plane magnetization of the island. (b) Constant-height image acquired in the center of the island at $z=80$~pm with $V=-1$~mV (size: $1\times1$~nm$^2$). We converted the recorded current into the corresponding effective $B$. (c) $d^2I/dV^2$ spectra acquired with the tip positioned above a Co atom of the island. The tip was moved from $z=145$ pm to $z=70$~pm. For clarity, the spectra are displaced vertically from one another by $5\times10^{-3}$ $G_0$/mV. (d) Exchange field $B$ extracted from the spectra of panel (c) using Eq.~(\ref{ham}). (e) $d^2I/dV^2$ spectra acquired at $z=80$~pm above a top site [red dot in panel (b)] and above a hollow site [blue dot in panel (b)]. The image corrugation in panel (b) is calibrated using the exchange coupling of the two spectra in panel (e). (f) Ratio between the tunnel current above a top site and the tunnel current above a hollow site. (g) Constant-height image of the same area as in (b) also acquired at $z=80$~pm, but with lower tunnel bias ($-5$~mV).
\label{fig3}}
\end{figure}

The fit to the line shape using Eq.~(\ref{ham}) and a dynamical scattering model (solid red lines in Fig.~\ref{fig1}d)~\cite{Loth2010,Ternes2015} is highly satisfactory and provides quantitative values of the spin-asymmetry $\eta$ and the exchange field $B$ exerted by the Fe atom onto the Nc molecule assuming a gyromagnetic factor of $g=2$. We find that the exchange field is an exponential function of $z$ (Fig.~\ref{fig2}c), allowing us to exclude a magnetic dipolar interaction. Using our DFT-computed effective spin of $\langle S_\text{Fe}\rangle\approx1.7$ (Supplemental Sec. IV), the exchange coupling is $|J|\approx 0.9$~meV at the shortest probed Fe-Nc distances. The magnetic anisotropy $D$, which corresponds to the average position of the spin-split peaks and dips, remains constant with distance (Fig.~\ref{fig1}e). This indicates that the intramolecular structure of Nc is preserved on the tip apex~\cite{Heinrich2015,Ormaza2017a}. For the data presented in Fig.~\ref{fig1}d, we find a spin asymmetry of $\eta=32\%$ at the highest fields measured (Fig.~\ref{fig2}d), the data collected on an ensemble of different Nc-tips on different Fe atoms yielding a lower average value of $\eta=23\%$. The observed spin asymmetry can be used to estimate the spin polarization of the elastic electrons by inferring $\mathcal{T}_{\text{\text{t}},\uparrow}/ \mathcal{T}_{\text{t},\downarrow}=0.54$ (Supplemental Sec. V). Using $\eta=23\%$, we find $\mathcal{T}_{\text{\text{s}},\uparrow}/ \mathcal{T}_{\text{s},\downarrow}=0.34$, which leads to an elastic spin polarization of $-69\%$. This value is in fair agreement with the DFT spin polarization of $-84\%$ (Fig.~S5).

The sign of the exchange interaction between a magnetic atom and a magnetic tip apex is determined by the competition of direct and indirect interactions and may vary with tip-atom distance~\cite{Tao2009}. To gain insight into the exchange coupling between the Nc-tip and Fe atom, we computed with DFT the exchange energy defined as $E_{ex}=E_\text{P}-E_\text{AP}$ at various Nc-Fe distances by fully relaxing the junction geometry (Supplemental Sec. IV); $E_\text{P}$ ($E_\text{AP}$) is the total energy of the junction with the spin directions of Nc and Fe in parallel (antiparallel) alignment. To facilitate the comparison to the the experimental findings, we take $z=0$ as the center-to-center distance between the closest carbon atom of Nc and the Fe atom, which is approximately $250$ pm. The exchange interaction favors an antiparallel alignment of the two spins (Fig.~\ref{fig2}e)\textemdash the junction geometry remaining constant up to $50$ pm where a chemical bond then starts to form between Fe and Nc (Fig.~S1c). The energy difference between antiparallel and parallel alignment is $13$ meV at $120$ pm, while no difference could be evidenced above $300$~pm. The antiferromagnetic coupling is short ranged and attributed to the direct hybridization of the Fe $d$-orbitals with the frontier molecular orbitals of nickelocene. 

In the following, we extend the proof-of-concept for the Nc-tip to a collection of atoms. In particular, we investigate a prototypical ferromagnetic surface consisting of a nanoscale cobalt island grown on Cu(111) (Fig.~\ref{fig3}a)~\cite{Diekhoner2003,Pietzsch2004,Pietzsch2006,Rastei2007}. The islands are triangular-like and two-layers high with typical lateral extensions $\ge10$~nm for the cobalt-coverage used (Supplemental Sec. I). They possess, at low temperature, an out-of-plane magnetization perpendicular to the copper surface (Inset of Fig.~\ref{fig3}a) ~\cite{Huang1994,Pietzsch2004,Oka2010}. Unlike the iron atom, the cobalt spin is fixed and we assume it to be $\Uparrow$ without loss of generality. Nickelocene adsorbs preferentially on cobalt, either on top of the nanoislands or on the bottom edge of the island as remarked for other molecules~\cite{Iacovita2008}. Nc-tips were routinely prepared by transferring a molecule from the edge of the island to the Cu-tip apex. Given the low molecular coverage, large pristine areas of cobalt may be found on the sample.

\begin{figure}
  \includegraphics[width=0.75\textwidth,clip]{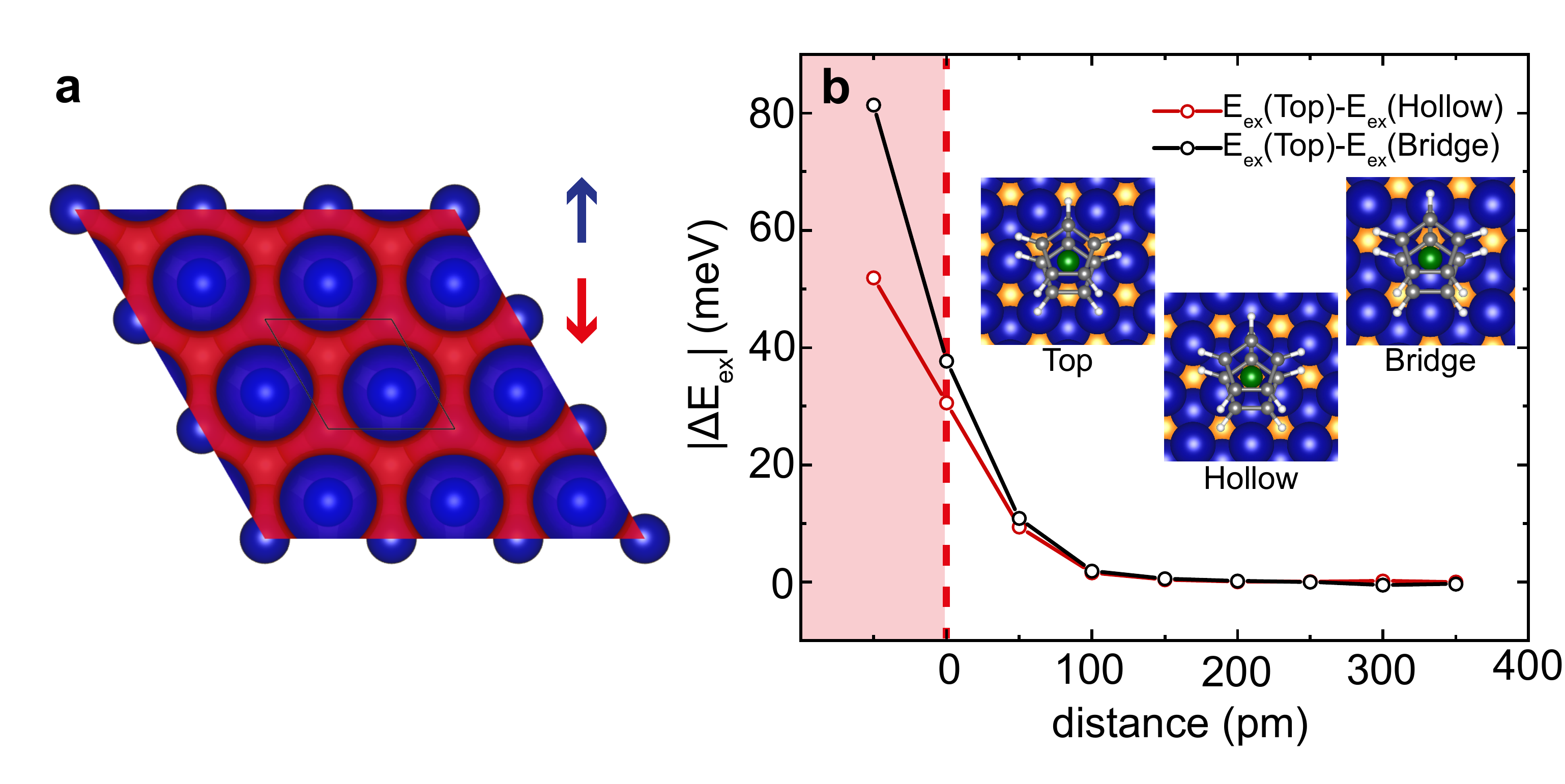}
  \caption{\textbf{Computed exchange interaction between nickelocene and cobalt.} (a) Isosurface of the spin density for a cobalt bilayer. Spin-down ($\downarrow$) and spin-up ($\uparrow$) densities are plotted in red and blue, respectively. Spin-up $d$ electrons are mainly located on the Co atoms, while spin-down $sp$ electrons are dispersive in nature~\cite{Diekhoner2003,Pietzsch2006,Oka2010}. The cobalt atoms have a magnetic moment of $1.76$~$\mu_\text{B}$, which is mainly carried by the $d$ orbitals ($d$: $1.81$~$\mu_\text{B}$, $p$: $-0.04$~$\mu_\text{B}$, $s$: $-0.01$~$\mu_\text{B}$). The spatial confinement of $sp$ electrons within the island~\cite{Diekhoner2003,Pietzsch2006,Oka2010} is not accounted for. (b) Difference in exchange energy ($|\Delta E_{ex}|$) between the top and the hollow positions (red), and between the top and bridge positions (black). The energy difference was computed as a function of tip distance to the cobalt surface. The filled red area indicates the contact regime where the Nc molecule is covalently bond to the surface. Inset: Junction geometry used in the DFT calculations defining the top, hollow, and bridge site.
\label{fig4}}
\end{figure}

In Fig.~\ref{fig3}b, we present a typical constant-height image acquired in a small area located in the center of a cobalt island (white square in Fig.~\ref{fig3}a) at a bias $V=-1$~mV, very close to the Fermi energy. The cobalt atoms of the island can be readily visualized with the Nc-tip, while this resolution is lost above non-magnetic Cu(111). This points to the magnetic origin of the contrast. Insight may be gathered through the vertical dependence of the spin excitation spectra acquired with the Nc-tip (Fig.~\ref{fig3}c). At distances $z=145$~pm above the cobalt surface, the spin excitation spectrum is similar to the spectra of Fig.~\ref{fig1}b, which indicates that also the island is spectroscopically dark. Interestingly, upon vertically approaching a Co atom in the cobalt island, the peak and dip in the $d^2I/dV^2$ spectrum progressively split apart. Using the spin Hamiltonian of Eq.~(\ref{ham}) and $g=2$, we find that the exchange field varies also here exponentially reaching values as high as $25$ T at the shortest distances explored ($z=70$~pm, Fig.~\ref{fig3}d). The exchange field originates mostly by direct orbital overlap and corresponds to an exchange coupling of $|J|\approx3.2$~meV taking our DFT-computed value of $\langle S_\text{Co}\rangle\approx0.9$ for the effective spin of a cobalt atom. The spectra show weak spin asymmetry ($\eta<5\%$), \textemdash in stark contrast to the single Fe atom, pointing to $\mathcal{T}_{\text{t},\uparrow}\mathcal{T}_{\text{s},\downarrow}\approx \mathcal{T}_{\text{s},\uparrow} \mathcal{T}_{\text{t},\downarrow}$. This reflects a compensation of the tip and sample spin-dependent transmissions. Taking as previously  $\mathcal{T}_{\text{\text{t}},\uparrow}/ \mathcal{T}_{\text{t},\downarrow}=0.54$, we find a $-55\%$ spin polarization for the elastic transmission of the island, which we attribute to spin-down $sp$ electrons in agreement with STM measurements carried out with a superconducting tip~\cite{Eltschka2017} \textemdash note that we have chosen the $sp$ electrons of the island to be spin down (Fig.~\ref{fig4}a). 

To clarify the observed atomic resolution in the constant-height images taken at bias voltages close to the Fermi energy, we compare in Fig.~\ref{fig3}e two spectra, one with the Nc-tip positioned above a Co atom (designated hereafter as a top site of the surface; red dot in Fig.~\ref{fig3}b) and the other with the Nc-tip above a hollow site of the surface (blue dot in Fig.~\ref{fig3}b). As shown, the exchange field varies among the two sites and with it the position of the low-energy excitation peak and dip. These move towards zero bias when the exchange field increases, shifting instead towards $e|V|=D$ when the exchange field decreases. The shift of the low-energy peak (dip) translates into a variation of the tunneling current $I$ by working at biases sufficiently low. In particular at $|V|=1$~mV and $z=80$~pm, we have a situation where the inelastic excitation to the first excited state is possible only when the Nc-tip is placed above the top site, while above the hollow site this channel is closed. This yields then a ratio of $I_{\text{top}}/I_{\text{hollow}}\approx 3$ (Fig.~\ref{fig3}f) producing a contrast in the constant-height image. At increased absolute bias ($|V|>5$~mV), $I_{\text{top}}/I_{\text{hollow}}$ is reduced to $1.25$ leading to a loss of contrast in the constant-height image (Fig.~\ref{fig3}g). To first approximation, the low-bias image shown in Fig.~\ref{fig3}b therefore reflects the spatial dependence of the exchange field at the atomic scale. Alternatively, it could also be possible to plot the exchange field from the spectra as a function of position, at the expense of considerably increasing the acquisition time of the image.

To confirm the magnetic origin of the contrast, we computed with DFT the exchange energy $E_{ex}=E_\text{P}-E_\text{AP}$ by varying the distance between the Nc-tip and the cobalt surface (Fig.~\ref{fig4}a and Supplemental Sec. IV). Three locations were investigated, corresponding to a Nc-tip laterally positioned above the surface with its nickel atom centered above a top, hollow and bridge site of the surface (Inset of Fig.~\ref{fig4}b). Just prior to the contact formation between Nc and the surface, which is the distance interval explored in the experiment, the exchange energy is markedly different between these sites (Fig.~\ref{fig4}b), because the local character of the $d$-electrons starts imprinting a lateral corrugation to the interaction. The exchange field can then be expected to change when moving the Nc-tip above the surface, in qualitative agreement with our experimental findings of Fig.~\ref{fig3}b. We stress that for a quantitative comparison, which is beyond the scope of the present study, the non-collinearity among magnetic moments of Co and Nc should be taken into account. 

To summarize, the spin excitation spectrum of a nickelocene molecule attached to the apex of a STM tip can be used to measure the magnetic exchange field of surface-supported objects. This is particularly dramatic for the case of a dense-packed magnetic layer, where the exchange field leads to a magnetic contrast with atomic-scale resolution. The spin excitation spectrum also allows for probing the spin asymmetry of these objects. Unlike conventional spin-polarized STM, the spin polarization of the 
object under investigation may then be determined with minimal influence of the probe on the system owing to the well-characterized molecular tip apex. In other words, this detection scheme offers to simultaneously obtain a magnetic contrast in two ways: by recording the exchange field across the vacuum gap as in pioneering magnetic exchange force microscopy experiments~\cite{Kaiser2007,Grenz2017}, and by measuring the spin polarization of the sample as in spin-polarized STM. A large variety of magnetic systems may be investigated that include model systems such as single atoms, surfaces with complex magnetic structures, or organometallic molecules on magnetic surfaces. The visualization of complex spin textures should benefit from an external magnetic field, making it even possible to experimentally determine the sign of the magnetic exchange interaction~\cite{Muenks2017}. A minor drawback of the technique is the presence of a tip-related molecular pattern in the images \textemdash that may however be corrected owing to the knowledge of the tip status, and the possible back-action exerted by the Nc-tip on the sample.

\begin{acknowledgments}
This work was supported by the Agence Nationale de la Recherche (Grant No. ANR-13-BS10-0016, ANR-15-CE09-0017, ANR-11-LABX-0058 NIE, ANR-10-LABX-0026 CSC). MT acknowledges support by the Heisenberg Program (Grant No.\ TE 833/2-1) of the German Research Foundation. NL acknowledges support by the Spanish MINECO (Grant No. MAT2015-66888-C3-2-R).
\end{acknowledgments}

%

\end{document}